\documentclass[aps,prl,twocolumn]{revtex4-2}
\usepackage{amssymb}
\usepackage{amsmath,bm}
\usepackage{graphicx,color}
 \usepackage{epstopdf}
\usepackage{epsfig}
\usepackage{undertilde}
\usepackage{multirow}
\usepackage[cal=pxtx]{mathalfa}
\usepackage{accents}
\DeclareMathAccent{\wtilde}{\mathord}{largesymbols}{"65}

\usepackage{stackengine}
\stackMath
\newcommand\tenq[2][1]{%
	\def\useanchorwidth{T}%
	\ifnum#1>1%
	\stackunder[0pt]{\tenq[\numexpr#1-1\relax]{#2}}{\scriptscriptstyle\sim}%
	\else%
	\stackunder[1pt]{#2}{\scriptscriptstyle\sim}%
	\fi%
}

\renewcommand{\vec}[1]{{\underline{#1}}}
\newcommand{\B}[1]{{\underline{#1}}} % XXX this is useful to make complex tensor homogeneous XXX

\begin{document}

\title{Constitutive  relations for colloidal gel}
\author{Saikat Roy$^{1}$}
\email{Corresponding authors: saikat.roy@iitrpr.ac.in}
\author{Yezaz Ahmed Gadi Man$^1$}
\affiliation{$^1$ Department of Chemical Engineering, Indian Institute of Technology Ropar, Rupnagar, Punjab, India 140001}

\begin{abstract}
The theoretical treatment of depletion gels with central interactions often involves expanding the free energy around a stress-free reference state to derive a constitutive relation between global stress and strain. The premise upon which the previous continuum theories are based, i.e., the stress-free reference state and the affine deformation, both of which do not hold in the context of amorphous gel materials. Gels never reach a true global minimum in the potential energy landscape and contain local regions of significant compressive and tensile stress, interspersed with zero-stress regions. Hence, expansion of free energy around a stressed reference state will produce scalar terms in harmonic expansion, the effects of which are qualitatively different from the terms appearing in the expansion around an unstressed reference state. In this study, we demonstrate the limitations of traditional continuum theories and propose simple constitutive relations that better capture the mechanical response of gel materials. The robustness of the proposed relations is established through large-scale numerical simulations of depletion and frictional gels across a vast parameter space.

\end{abstract}

\maketitle
\textit{Introduction}: Colloidal suspensions can give rise to diverse rheological responses based on the external driving, the nature of the interaction potential, and the distance from the gel point. Dilute suspensions with sufficiently strong short-ranged attractions form a space-spanning, rigid, percolated network that can sustain external deformation up to a critical strain. The metastable structure at the gel point exhibits significant spatial heterogeneity in the arrangement of its constituents, owing to the presence of large voids. Moreover, the thermal fluctuations in the colloidal gel system freeze as soon as the gel network forms, resulting in a spatially inhomogeneous quenched stress field from the outset\cite{nampoothiri2022tensor,vinutha2023stress}. The interplay between the heterogeneous gel structure and the inherent internal stress field makes the strain field anything but affine under a global macroscopic strain. 
%The gel reaches an attractive glassy state at a higher volume fraction ($\phi>0.5$) with a spatially homogeneous structure. 

The diverse mechanical responses observed across different volume fractions and attraction strengths were elegantly reconciled in a pivotal work by \citet{zaccone2009elasticity}. In this work, the free energy of the depletion gel system is essentially expanded around an unstressed reference state on an appropriate length scale, $\xi$. Under the assumption of affine deformation, the expression for the shear modulus becomes \cite{zaccone2009elasticity}:
\begin{equation}\label{shearalessio}
 G_c\approx(2/5)\pi^{-1}\phi_cz_c\tilde{\kappa}\xi^{2-d}
 \end{equation}
where $G_c$, $\phi_c$, $z_c$ represent the cluster level shear modulus, volume fraction, and mean coordination number, respectively. $\tilde{\kappa}$ and $\xi$ denote the cluster-level bond stiffness and the effective cluster size. In essence, the complex effect of spatial structural heterogeneities on the mechanical response is conjectured to be captured by a characteristic length scale, whose variation with attraction strength and volume fraction manifests varied rheological behavior. Conversely, the previous experimental and simulation observations \cite{ramakrishnan2004elasticity,whitaker2019colloidal} on the depletion gels imply a very weak dependence of $\xi$ on the volume fraction and attraction strength. Hence, the characteristic length scale proves inefficient at capturing the varied elastic response. The bond stiffness typically sets the scale of the cluster-level stiffness for small clusters, and prior measurements of bond stiffness also show little to no change with attraction strength and volume fraction. The only remaining strong candidate to explain the variations is the cluster connectivity. Identifying rigid, elastically active clusters within the gel network is challenging. The l-balanced graph partitioning \cite{whitaker2019colloidal} was employed to decompose the gel network into clusters, but the rigidness of the identified clusters was not rigorously proven.
The clusters were deemed rigid based on the mean coordination number exceeding the minimum isostatic contact number required for stability in the central force network. However, in reality, a dense cluster can rotate freely under external deformation without contributing to the network's elastic energy. Hence, mere constraint-counting arguments fail to provide an accurate description of the rigidity. A careful inspection of the previous experimental observations \cite{whitaker2019colloidal} hint towards a possible discrepancy between the measured data and the theoretical prediction \cite{zaccone2009elasticity}: upon increasing the dimensionless attraction strength, a sevenfold rise is observed for the shear modulus whereas the product of the cluster number density, $N_c$, and coordination number, $z_c$ displays only a threefold rise. \\
Gels also exhibit intriguing power law growth of stress with volume fraction 
\cite{roy2016yieldinga,roy2016yieldingb,roy2016universality,roy2020micro,buscall1987consolidation} under another interesting and practical mode of deformation, i.e., uniaxial consolidation, which is widely employed in the industry for fabrication of ceramic materials, cosmetics, paints, drilling,  fuel cells, and drying of colloidal dispersions. Very recent investigations on depletion gels with central interactions \cite {tateno2024mechanical,tateno2025void,gadi2025origin} and the gels with frictional non-central interactions \cite{gadi2023micro,islam2021normal} demonstrate the presence of significant normal stress difference (NSD) during consolidation and a power law scaling of compressive yield stress. These observations defy all previous models in the literature, which employ fractal scaling arguments without a microscopic basis. The gels formed in the presence of frictional interactions are fractal in nature in the beginning but lose fractal correlations in the intermediate to high volume fractions \cite{seto2013compressive}. On the other hand, the depletion gels, which are not fractal, still exhibit power-law scaling. In this letter, we convincingly demonstrate that the previous theoretical formalism \cite{zaccone2009elasticity} fails miserably to predict the power-law growth of stress and the evolution of NSD. We also propose simple constitutive relations that relate the components of the stress tensor to the appropriate internal state parameters, correctly encoding the non-linear effects of the quenched stresses. The generality of the proposed relations is rigorously tested through large-scale numerical simulations of frictional and depletion gels undergoing quasistatic uniaxial compression. \\ 
\textit{Methodology}: 
The simulation begins with the random placement of non-overlapping $N=20000$ bi-disperse (1:1.4 size ratio) spheres in a three-dimensional simulation box of size $L_x=L_y=80a, L_z=257a$, where $a$ denotes the radius of the smallest particle. Quasistatic uniaxial compression is enforced by the slow axial ($\hat{\mathbf{e}}_z$) movement of the top and bottom walls at a constant speed. The periodic boundary condition is employed on the transverse directions ($\hat{\mathbf{e}}_x$, $\hat{\mathbf{e}}_y$). The hydrodynamic forces are negligible compared to the dominant elastic force in the limit of small strain rates and are therefore not considered in the simulation. For the depletion gel, the pairwise interaction is modeled using the Morse potential, with a very short attraction range corresponding to the experimental range of non-adsorbing polymer depletants. The dimensionless strength of the potential minimum, $\beta U_0$, is set at $50$ and $100$ to mimic strong gels\cite{dibble2006structure}, where $\beta=1/k_BT$ ($k_B$ being the Boltzmann constant). \\
The frictional gel simulation incorporates non-hydrodynamic contact friction, and the interparticle interaction is modeled in the spirit of traditional discrete element method (DEM) simulations employed for granular materials \cite{cundall1979discrete}. The particles are given random thermal kicks to the translation degrees of freedom, compensated by a viscous damping term. The stochastic Langevin equation is solved to update the particle position and velocities. The gel is slowly compressed to various target volume fractions ($\phi=0.25-0.5$) and then relaxed to ensure mechanical equilibrium. For details of the simulation, see the supplemental material\cite{SM}.\\
\textit{Failure of previous continuum theories}: 
Given the tensorial structure of the previous constitutive relations \cite{zaccone2009elasticity} and its widespread use in predicting the shear modulus of the depletion gel system, the applicability of the theory to other deformation scenarios is naturally expected. Hence, we subject the continuum theory to predict the salient features of the uniaxial compression, such as the evolution of NSD\cite{tateno2024mechanical,tateno2025void} with $\phi$. Upon assuming affine deformation,  the Cauchy-Born expansion of free energy around an unstressed reference state on an appropriate length scale ($\xi$) relates the global stress, $\sigma_{ij}$, to the macroscopic strain via the following equation\cite{zaccone2009elasticity}:  
\begin{equation}\label{alessionstressstrain}
\sigma_{ij}=\frac{\tilde{\kappa}N_c z_c}{2}\langle \xi^2n_in_jn_kn_l \rangle e_{kl}
 \end{equation}
where $\langle..\rangle$ represents orientational averaging, $n_i$ denotes the Cartesian components of the cluster center to center normal unit vector. Note that $N_c$ is related to the cluster level volume fraction via $\phi_c=N_c\pi\xi^3/6$. Under the assumption of no correlations between $\xi$ and $n_i$ and isotropy in the cluster orientations, the expression for $\sigma_{ij}$ simplifies to: 
\begin{equation}\label{alessionstressstraincon}
\sigma_{ij}=\frac{\tilde{\kappa}N_c z_c}{30} \,\xi^2 \left(e_{ll}\delta_{ij}+2e_{ij}\right)
\end{equation}
The global strain tensor for an uniaxial compression in $\hat{\mathbf{e}}_z$ direction becomes $e_{ij}=e_0\delta_{zi}\delta_{jz}$. Plugging this into Eq.~\ref{alessionstressstraincon}, gives the axial ($\sigma_{zz}$) and transverse stress components ($\sigma_{xx}$, $\sigma_{yy}$):
\begin{equation}\label{stresscomp}
\sigma_{zz}=\frac{\tilde{\kappa}N_c z_c}{10} \,\xi^2 e_0;\,
\sigma_{xx}=\sigma_{yy}=\frac{\tilde{\kappa}N_c z_c}{30} \,\xi^2 e_0;\,
%\sigma_{yy}=\frac{\tilde{\kappa}N_c z_c}{30} \,\xi^2 e_0;\,
\end{equation}
Two normal stress differences can be defined in $\textit{three dimensions}$: $N1=(\sigma_{zz}-\sigma_{xx})/\sigma_{zz}$ and $N2=(\sigma_{zz}-\sigma_{yy})/\sigma_{zz}$. From Eq.~\ref{stresscomp}, the normal stress difference when normalized by the axial stress yields a constant value $2/3$ for both $N1$ and $N2$, which contradicts the experimental evolution of NSD.
Very recent experiments on depletion gel systems \cite{tateno2024mechanical,tateno2025void} present significant NSD at lower $\phi$, which reduces as $\phi$ approaches the glassy state. It is important to emphasize that the theoretical expression for the non-dimensional NSD neither contains any cluster connectivity ($z_c$) term nor any $\phi_c$-dependent term nor any length scale ($\xi$) nor any cluster level stiffness term ($\tilde{\kappa}$) whose evolution could have captured the observed NSD evolution with $\phi$. The failure of the previous continuum theories originates from the two rigid assumptions: i) the stress-free reference state, ii) affine deformations, both of which fail outright in the context of amorphous gel systems. To resolve the conundrum regarding the evolution of NSD and the power law growth of stress during the uniaxial compression of gels, we propose simple constitutive relations that relate the macroscopic response to the microscopic features of the network. \\
\textit{Constitutive relations}: The global stress tensor for the gel system under static equilibrium can generically be expressed\cite{landau1986theory} in terms of volume averages of dyadic products between the pairwise interaction force ($f_j^{c}$) and the branch vector ($l_i^{c}$):
\begin{equation}\label{stressgeneral}
\sigma_{ij}=\frac{n_c}{2}\langle l_i^c f_j^c \rangle
\end{equation}
where $n_c$ represents the contact number density and $i$, $j$ refer to the Cartesian components as per the usual index notation. Assuming statistical independence between $f_j^{c}$ and $l_i^{c}$, the stress tensor can be conveniently evaluated by a directional average over orientations ($\hat{\mathbf{n}}$) in the limit of large system size:
\begin{equation}\label{stressorient}
\sigma_{ij}
=
\frac{n_c}{2}
\int
\overline{l_i}(\hat{\mathbf n})\,
\overline{f_j}(\hat{\mathbf n})\,
P(\hat{\mathbf n})\, d\hat{\mathbf n}.
\end{equation}
where $P(\hat{\mathbf n})$ is the probability density function of the contact orientations. Since $P(\hat{\mathbf n})=P(-\hat{\mathbf n})$, it can be approximated by a scalar valued smooth function of $\hat{\mathbf n}$ up to second order:  $G(\hat{\mathbf n})=H+H_{ij}n_in_j$. Upon applying the least square approximation\cite{ken1984distribution}, the polynomial expansion can equivalently be expressed as spherical harmonic expansions: $P(\theta,\phi)=\sum_{\ell=0}^{\infty}\sum_{m=-\ell}^{\ell}A_{\ell}^{m}\,Y_{\ell}^{m}(\theta,\phi)$ in $\textit{three dimensions}$. The distribution of orientations exhibits azimuthal symmetry as observed in the simulation. To respect the $\pi$-periodicity and azimuthal symmetry, the only admissible spherical basis functions up to second order are $Y_0^{0}=1$ and $Y_2^{0}=3\cos^2\theta -1$. Upon applying the normalization, $\int P(\theta,\phi)\,\sin\theta\, d\theta\, d\phi=1$, the orientation probability density function is represented as:
\begin{equation}\label{contactorient}
P(\theta,\phi)\approx P(\theta)\approx \frac{1}{4\pi}\left\lbrace 1+a(3\cos^2\theta-1) \right\rbrace
\end{equation}     
where $a$ denotes the anisotropy in the contact orientations. The branch vector, $l_i$ for a bidisperse population of particles, can be represented as $\bar{D}n_i$, where $\bar{D}$ is the mean particle diameter. In the presence of friction, the interaction force in the local reference frame is given as $\overline{f_j}(\hat{\mathbf n})=\overline{f_n}(\hat{\mathbf n})n_j+\overline{f_t}(\hat{\mathbf n})t_j$, where $\overline{f_n}(\hat{\mathbf n})$ and $\overline{f_t}(\hat{\mathbf n})$ denote the average normal and tangential components of the interaction force for a group of contacts with orientation $\hat{\mathbf n}$. For depletion interaction, only $\overline{f_n}(\hat{\mathbf n})$ survives.  Similar to $P(\hat{\mathbf n})$, $\overline{f_n}(\hat{\mathbf n})$ being an even function of $\hat{\mathbf n}$ can be expressed as:
  \begin{equation}\label{fn}
\overline{f_n}(\theta,\phi) \approx \overline{f_n}(\theta)\approx \overline{f_m}\left\lbrace 1+a_n(3\cos^2\theta-1) \right\rbrace
\end{equation} 
where $\overline{f_m}$ is the mean normal force over all contacts, $a_n$ denotes the anisotropy in the directional variations of normal forces. The azimuthal symmetry is also preserved for the normal forces as implied by the simulation observations, hence no $\phi$ dependent term is incorporated in the expansion. In the absence of an external rotational field and twisting friction at the contact scale, the azimuthal component of the tangential force is expected to be negligible, which is also supported by a very negligible background value observed in the simulation.  The tangential force then reduces to $\overline{f_{t\theta}}(\theta)\,\hat{\mathbf e}_{\theta}$ for azimuthal symmetry. It can be expanded in the vector spherical harmonics respecting the azimuthal symmetry: $\overline{f_{t\theta}}(\theta)\,\hat{\mathbf e}_{\theta}=\sum A_{l}\frac{dP_l(\cos\theta)}{d\theta}\hat{\mathbf e}_{\theta}$, where $P_l(cos\theta)$ are the Legendre polynomials. In view of the symmetry requirement $\overline{f_{t\theta}}(-\hat{\mathbf n})=-\overline{f_{t\theta}}(\hat{\mathbf n})$, the lowest order expansion that captures the orientational structure of the tangential forces is $\overline{f_{t\theta}}(\theta)\,\hat{\mathbf e}_{\theta}=-\overline{f_m}a_t\,\sin2\theta\,\hat{\mathbf e}_{\theta}$. $a_t$ defines the anisotropy in the tangential forces.
\begin{figure}[htbp!]
\includegraphics[scale=0.28]{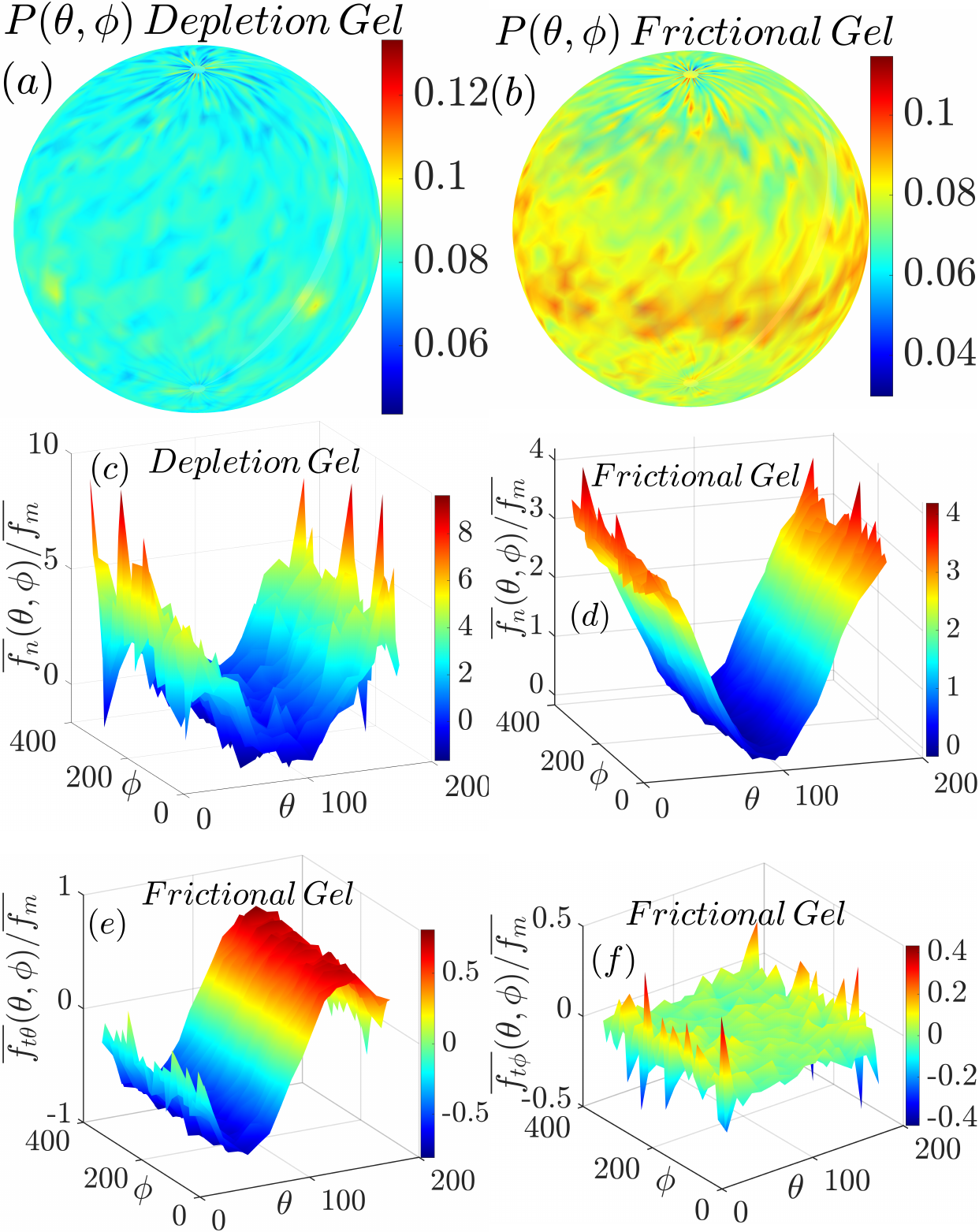}
\caption{(a) Contact orientation distribution is plotted on a unit sphere for the depletion gel at $\phi=0.25$ for $\beta U_0=50$. (b) The same distribution is plotted for the frictional gel at identical conditions. Isotropy in contact orientations is also maintained at higher values of attraction strengths and volume fractions (see \cite{SM}). (c)-(d) show the universality in angular modulation of the normal forces irrespective of the frictional interactions or lack thereof. (e) depicts the azimuthally symmetric sinusoidal modulation of $\overline{f_{t\theta}}(\theta,\phi)$ (f) Negligible isotropic background value as observed for $\overline{f_{t\phi}}(\theta,\phi)$.}
\label{anisotropy}
\end{figure}
The expressions for the axial ($\sigma_{zz}$), transverse stress components ($\sigma_{xx}\,,\sigma_{yy}$) and subsequently the dimensionless normal stress differences ($N1, N2$) can easily be obtained by substituting $P(\theta)$, $\overline{f_n}(\theta)$, and $\overline{f_{t\theta}}(\theta)\,\hat{\mathbf e}_{\theta}$ into Eq.~\ref{stressorient} and integrating over the solid angle:
\begin{align}
&\sigma_{xx} = \sigma_{yy}=\frac{3\mathcal{P}}{4}\left(-\frac{8}{15}(a_n+a_t)+\frac{4}{3}\right)\label{transversestress}&\\
&\sigma_{zz} = \frac{3\mathcal{P}}{4}\left(\frac{16}{15}(a_n+a_t)+\frac{4}{3}\right)\label{axialstress}&\\
&N1=N2=\frac{ \frac{2}{5}\left(a_n+a_t\right)}{\left(\frac{1}{3}+\frac{4}{15}(a_n+a_t)\right)}\label{nsd2}&
\end{align} 
where, $\mathcal{P}=\sigma_{ii}/3$ is the system pressure and is given as $\phi z \overline{f_m}/\pi \bar{D}^2 $.
The calculation of the contact and force anisotropy parameters from the discrete simulation data is readily carried out by defining three second order tensors: $\Phi_{ij}\approx\frac{1}{N_c}\sum n_i n_j$, $\Xi_{ij}\approx\frac{1}{N_c}\sum \overline{f_n} n_i n_j$ and $\chi_{ij}\approx\frac{1}{N_c}\sum \overline{f_t} n_i t_j$, where $N_c$ denotes the total number of contacts. The anisotropy paramters are then given as: $a=\frac{5}{2}\frac{(\Phi_{max}-\Phi_{min})}{\Phi_{ii}}$, $a_n=\frac{5}{2}\frac{(\Xi_{max}-\Xi_{min})}{\Xi_{ii}}$, and $a_t=\frac{5}{2}\frac{(\chi_{max}-\chi_{min})}{\Xi_{ii}}$, where $ max$ and $min$ subscripts represent the maximum and minimum principal values of the respective tensors. 
\begin{figure}[htbp!]
\includegraphics[scale=0.13]{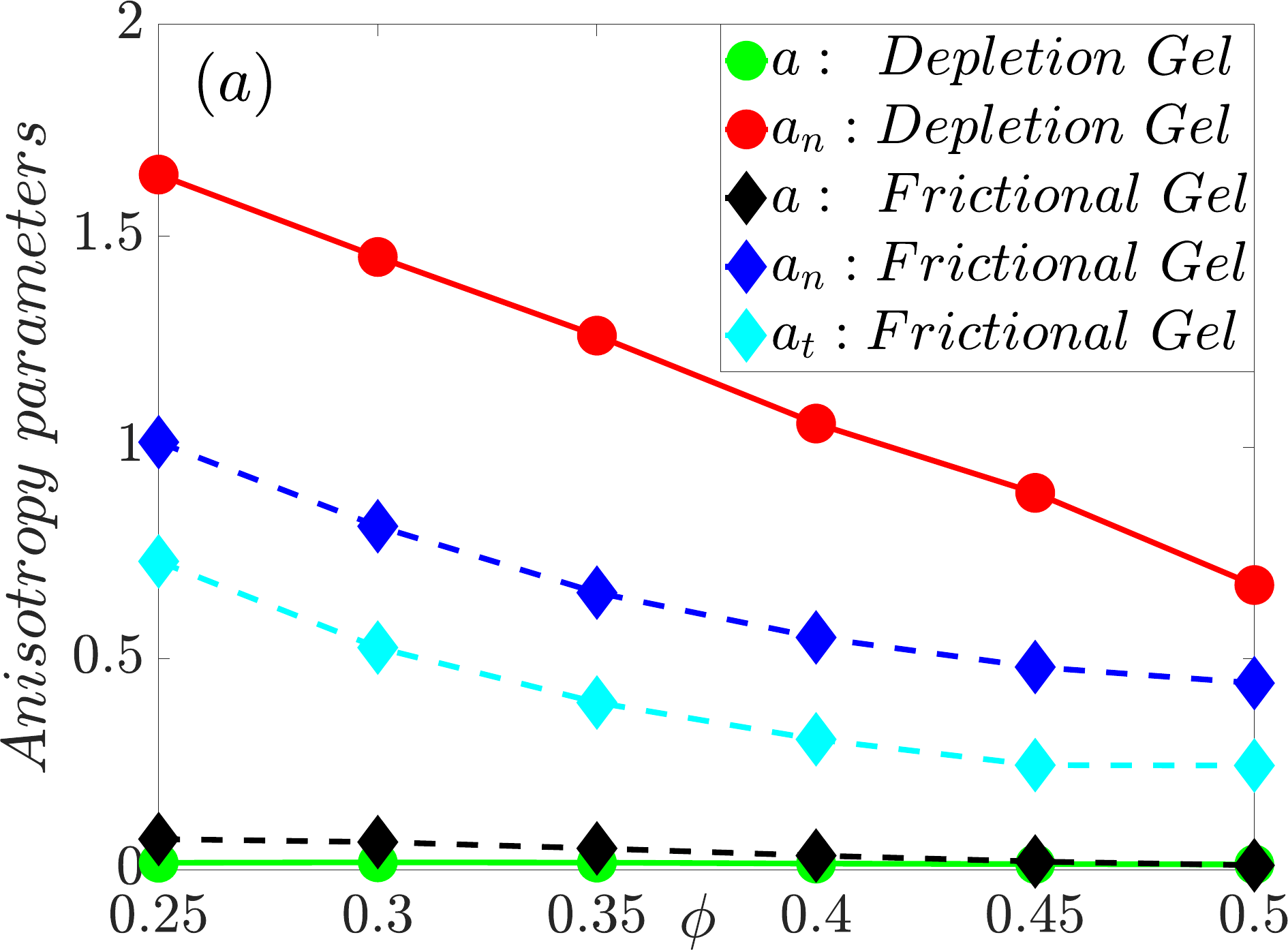}
\includegraphics[scale=0.13]{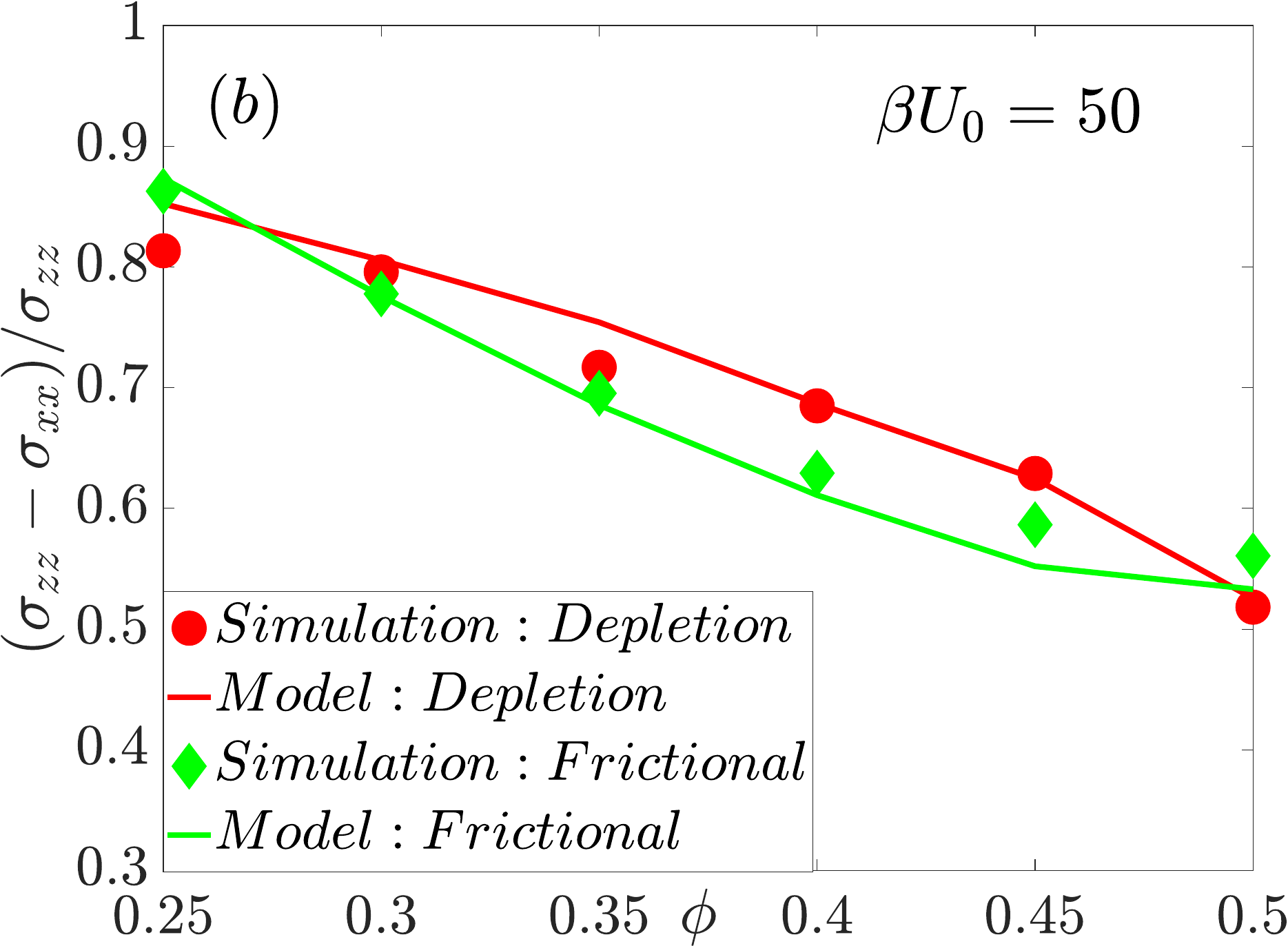}
\caption{(a) Evolution of contact anisotropy, $a$, and normal force anisotropy parameter, $a_n$, is shown as a function of volume fraction for gels with central and non-central interactions. In addition, the evolution of the tangential force anisotropy, $a_t$, is presented for the frictional gel. Here, $\beta U_0=50$. Similar values are observed for higher attraction strengths. (b) Comparison of dimensionless NSD values as obtained in the simulation with those of the model prediction. Here, $\beta U_0=50$. Model comparison is also remarkable for a higher $\beta U_0$ (not shown).  }
\label{anatNSD}
\end{figure}
\begin{figure}[htbp!]
\includegraphics[scale=0.4]{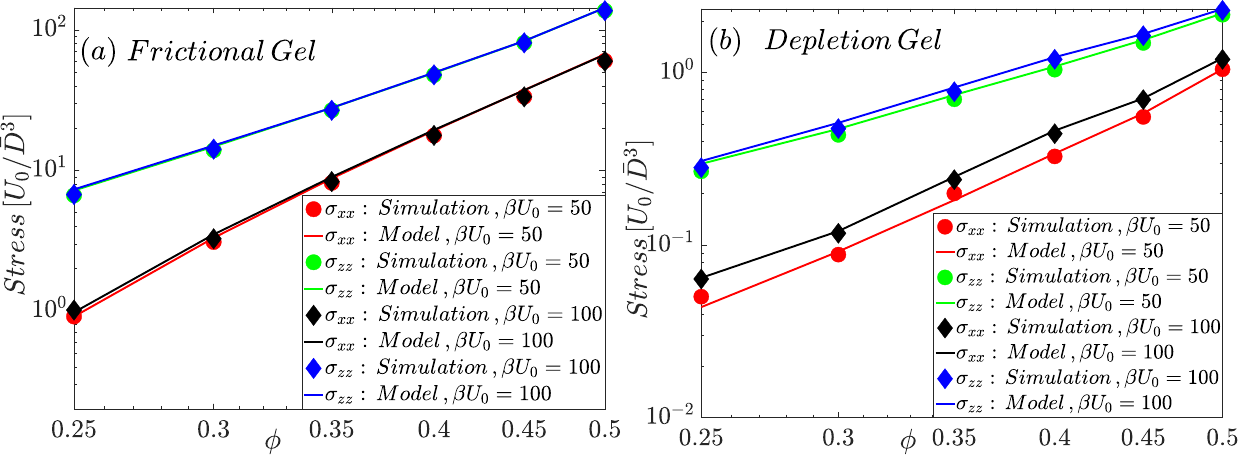}
\caption{Comparison between the prediction of the proposed model and the simulation observations for the relevant stress components covering a wide range of volume fractions, attraction strengths, and the nature of interparticle interactions. }
\label{stressNSD}
\end{figure}

\textit{Contact and force network anisotropy}: 
To understand the role of contact orientations on supporting the global stress, we plot the orientational probability density function, $P(\theta,\phi)$, in Fig.~\ref{anisotropy}(a) and (b) for both the depletion and frictional gels. Intriguingly, the gel contact network maintains an almost isotropic structure throughout the consolidation process\cite{tateno2024mechanical}, with a slight tendency to proliferate contacts at $\theta=\pi/2$ in the presence of friction. The isotropy in the orientations can also be quantitatively confirmed by the minimal value observed for the contact anisotropy parameter, $a$ (see Fig.~\ref{anatNSD} (a)). Hence, $a$ has been set to zero to represent an isotropic $P(\theta,\phi)$ given by $1/4\pi$. The persistence of isotropy even under an anisotropic driving implies very little rearrangement of the contact network held tightly by the strong attraction. In contrast, contact anisotropy is significant in the dry granular media and makes a notable contribution to the evolution of the global stress. In the absence of cohesive interactions, granular particles can easily roll or slide, leading the fabric to evolve under an incompatible load.\\
Next, the pertinent question is: What is the origin of stress anisotropy in a perfectly isotropic gel contact network? To answer this, it is imperative to investigate the structure of the force network that percolates through the isotropic contact network. Fig.~\ref{anisotropy}(c) and (d) show the directional variations of the average normal force for two classes of gels. $\overline{f_n}(\theta,\phi)$ presents azimuthal symmetry and a universal form dictated by the spherical harmonic expansion as per Eq.~\ref{fn}, irrespective of the interaction potential and volume fraction (see also \cite{SM}). Unlike the contact network, the normal force network exhibits significant anisotropy as evidenced by the appearance of larger forces along the compression direction ($\theta=0$) compared to the transverse direction ($\theta=\pi/2$). For the frictional gel, the tangential force network is also relevant, and its angular modulation is plotted in Fig.~\ref{anisotropy} (e)-(f). The polar component of the tangential force respects the azimuthal symmetry and follows a sinusoidal modulation in $2\theta$ as predicted. The anisotropy in the tangential force network is also evident. Interestingly, the same orientational structure for $\overline{f_{t\theta}}(\theta,\phi)$ is observed for all volume fractions and attraction strengths\cite{SM}. The azimuthal component of the tangential force is found to be insignificant and appears as an isotropic background as expected. \\
To connect the anisotropic force network with the observed stress anisotropy, the strength of the anisotropy in the normal and tangential force network is determined, and its evolution with the volume fraction is plotted in Fig.~\ref{anatNSD}(a). In the dilute regime, the bulk of stress transmission occurs via anisotropic force chains aligned along the compression direction, giving rise to pronounced mechanical anisotropy. However, the anisotropic force chain network buckles with increasing $\phi$ and transfers part of the axial load to the transverse directions, leading to diminishing values for $a_n$ and $a_t$. The depletion gel exhibits more substantial force anisotropy than the frictional gel, which could be attributed to the absence of frictional interactions that facilitate stress homogenization. It is worth noting that the evolution of both the contact and force anisotropies is apparently insensitive to the strength of interparticle attraction, which hints towards a mechanically self-organized state \cite{tateno2024mechanical,tateno2025void} arising purely out of external driving. Moreover, the precedence of the diffusion-limited aggregation process over the gel's arrest in a metastable state renders the gel structure invariant at all length scales across different values of the attraction strength\cite{SM}. \\
\textit{Predictive capabilities of the proposed model}:
Finally, we test the constitutive relations as laid out in Eqs.\ref{transversestress}-\ref{nsd2} for their ability to capture the evolution of relevant stress components. In Fig.~\ref{stressNSD}(a)-(b), we show the comparison between the model prediction and the simulation data for the axial ($\sigma_{zz}$) and the transverse stress ($\sigma_{xx}$). The agreement is remarkable across both frictional and depletion gels at varied attraction strengths, demonstrating the generality and robustness of the proposed model. Interestingly, the axial stress when normalized by $U_0/\overline{D}^3$ shows an excellent collapse for different values of $\beta U_0$ with and without friction, implying the strength of attraction sets a natural stress scale in the system and accordingly dictates the scale of mean normal force, $\overline{f_m}$. Transverse stress also shows similar collapse for the frictional gel. However, for the gels with central interactions, $\sigma_{xx}$ presents a weaker scaling with $U_0$. \\
As a final rigorous test of the model, we subject it to predict the NSD evolution as observed in the simulation and experiments\cite{tateno2024mechanical} in Fig.~\ref{anatNSD}(b). Our model successfully captures the NSD at different stages of the consolidation process across a wide range of gels and varying attraction strengths. Here, it is worth noting that the expression for the normalized NSD (Eq.~\ref{nsd2}) contains only $a_n$ and $a_t$, which emphasize the important role of the force network anisotropy in modulating the global stress. None of the existing continuum models accounts for inhomogeneous stress transmission via an anisotropic force network, thereby failing to capture the NSD evolution.  \\
\textit{Outlook:} 
In summary, we convincingly demonstrate the inadequacies of the existing continuum theories that rely on the traditional stress-strain route to predict the gel's mechanical response. The absence of a zero-load reference state and the inhomogeneities in the initial internal stress field prevent the formulation of any meaningful constitutive relations between stress and strain. In this letter, we propose a fresh approach to incorporate the nonlinear effects of quenched stresses and the adaptation of the force skeletons by tracking the evolution of crucial internal state parameters, such as the mean normal force and the anisotropy of the contact and force networks. With the help of the scalar and vector spherical harmonic expansions, the directional data, such as contact orientations, normal and tangential forces, are represented in a tractable manner respecting the requisite symmetry of the scalar and vector-valued functions. Finally, we introduce a set of constitutive relations that relate the macroscopic stress tensor to the pertinent state parameters. The generality of the proposed relations is rigorously tested by performing large-scale numerical simulations for \textit{three dimensional} frictional and depletion gels. Also, our results strongly suggest that the mechanical response of the gel system is mainly governed by particle-length-scale phenomena, rather than the popularly held cluster-length-scale mechanism.\\
\textit{Acknowledgement-}S.R. acknowledges the support of SERB under Grant No. CRG/2023/003115.
\\
\section{Supplemental Material: Constitutive  relations for colloidal gel}

\section{Depletion gel simulation with central interaction force}
To obtain a microscopic understanding of the consolidation process in a strongly aggregating depletion gel, we carry out Langevin dynamics simulations using the open-source code Large-scale Atomic/Molecular Massively Parallel Simulator (LAMMPS) \cite{plimpton1995fast}. The spatiotemporal evolution of the colloidal gel system is dictated by the interplay between various forces, including stochastic thermal forces, pairwise interaction forces, long-range hydrodynamic forces, and forces resulting from external deformation. We employ very small strain rates ($\approx 10^{-4} s^{-1}$) to mimic quasistatic compression and to minimize the inertial effects. In the limit of slow deformation, the ratio of the hydrodynamic force ($\eta \dot{\epsilon}$) to the elastic force on a particle becomes negligible ($G\epsilon$). Here, $\eta$, $\dot{\epsilon}$, $G$, and $\epsilon$ denote solvent viscosity, strain rate, particle modulus, and strain, respectively. Previous experimental studies also demonstrate the dominance of network stress over hydrodynamic stress for the case of quasistatic deformation once a percolating network has formed. In addition, since we focus on the strong gel without any long-range Coulomb repulsion, the flocculation process is swift and not barrier-limited. In this case, the microstructure of the gel network freezes quickly, which inhibits any further rearrangement due to long-range hydrodynamic forces. Even though the hydrodynamics will play a role in the evolution of structure and age-dependent properties of weak gels, previous simulations \cite{varga2016hydrodynamic,de2019hydrodynamics} provide strong evidence that the microstructure of the gel network is very weakly dependent on the hydrodynamics in the limit of strong attraction strength. In light of the aforementioned experimental and simulation observations, as well as a simple order-of-magnitude analysis of the pertinent forces, the hydrodynamic forces are not considered in the simulation.\\
The constituents in the depletion gel system experience a short-range entropic attraction force induced by the non-adsorbing polymer depletants. The pairwise interaction is modeled using the Morse potential to capture both the short-range attraction and the nearly hard-sphere repulsion upon overlap. The higher order gel structure \cite{royall2021real} is better reproduced with the Morse potential than the traditional one-component Asakura and Oosawa (AO) interaction potential. The form of the potential employed in the simulation is given as,
 \begin{equation}\label{eq:1}
   \beta U(r) = 
    \begin{cases}
    -\beta U_0\;\{2e^{[-\kappa(r-r_0)]}-e^{[-2\kappa(r-r_0)]}\},& r < r_c \\
    0.& r \ge r_c
    \end{cases}
 \end{equation}
where, $\beta=1/k_BT$, $\beta U_0$ denotes the dimensionless strength of the potential minimum and $\kappa^{-1}$ dictates the range of attraction. $r_0=a_i+a_j$ represents the equilibrium separation distance between a particle pair, where $a_i$ and $a_j$ are the radii of the $i^{th}$ and the $j^{th}$ particles. $r_c$ sets the potential cutoff. A linear correction term is also introduced to the potential to ensure that both the potential energy and force become zero smoothly at the cutoff. The minimum of the potential is set at $\beta U_0=50$ in the simulation to mimic a strong gel\cite{dibble2006structure}. The attractive strength is also varied to see its effect on the mechanical response. The attraction range is kept very short ($\kappa a=21$), representative of an experimental range with non-adsorbing polymer depletants\cite{ap1999delayed,shah2003viscoelasticity,dibble2006structure}, where $a$ denotes the radius of the smallest particle.
$N=20000$ bi-disperse ($1:1.4$ size ratio) spheres are placed randomly in a three-dimensional simulation box of size $L_x=L_y=80a, L_z=257a$. Subsequently, energy minimization with a soft potential is employed to obtain a non-overlapping configuration. All length scales in the simulation are measured in terms of the smallest particle radius. The initial volume fraction of the colloidal system is $\phi\approx0.1$, which is well below the gelation point. Thermostatting is applied only to the translational degrees of freedom of the colloidal particles via random thermal kicks, compensated by a viscous damping term that mimics the effect of an implicit solvent\cite{gadi2025origin}. The stochastic Langevin equation is solved to update the particle position and velocities,
\begin{eqnarray}
m_i \frac{d^2 \B x_i}{dt^2} =\B F_i -m_i\eta_t \frac{d\B x_i}{dt} +\B f_i(t)
\end{eqnarray}
where $\B f_i(t)$ represents a $\delta$-correlated random force term having zero mean with the following properties:
\begin{equation}
\langle \B f_i(t) \cdot \B f_j(t+\tau)\rangle = 2 \Gamma \delta (\tau) \delta_{ij}
\end{equation}
Here $\eta_t$ denotes the translational damping coefficient, $m_i$ denotes the mass of the particle $i$, $\B F_i$ is the net force on particle $i$  due to all pairwise interactions, $\Gamma$ is the strength of fluctuations, and $\B x_i$ denotes the translational degrees of freedom. The interparticle interaction force due to the Morse potential is summed vectorially to yield the net force on a given particle $i$,
\begin{equation}
\B F_i=-\sum_j \frac{\partial U(r_{ij})}{\partial r_{ij}} \frac{\B r_{ij}}{|\B r_{ij}|}
\end{equation}

where, $U(r_{ij})$ represents the pairwise Morse potential between particle $i$ and $j$ separated by a center to center vector $\B r_{ij}$. 

The thermal fluctuations, along with short-range attractive forces, lead to the formation of smaller aggregates, which are brought closer together by uniaxial compression to form larger aggregates and a space-spanning, percolating rigid network at the gel point. Quasistatic uniaxial compression is effected by the slow axial ($\hat{\mathbf{e}}_z$) movement of the top and bottom walls at a constant speed, $V_0$. Walls consist of a monolayer of particles that interact with the system particles via the Morse potential, using the same parameters stated previously. The periodic boundary condition is enforced on the transverse directions ($\hat{\mathbf{e}}_x$, $\hat{\mathbf{e}}_y$ ). The gelation volume fraction and the corresponding microstructure in the simulation will depend on the two competing time scales: the Brownian diffusion time scale and the time scale set by the strain rate. For a fixed strength of stochastic fluctuations, a higher gelation concentration with dense microstructure can be attained by increasing the deformation rate.
In contrast, slow compression will give rise to an arrested state with a tenuous microstructure at a lower volume fraction. $V/V_0$ is set as $18.58$ in the simulation to capture the slow consolidation dynamics of the depletion gel system. $V$ is the characteristic velocity scale set by the thermal fluctuations and is given by $\sqrt{k_BT/m}$.   

Since the gel network is expected to explore the mechanically balanced inherent states under slow compression, the axial stress values obtained at a given volume fraction will correspond to the compressive yield stress values \cite{buscall1988scaling,buscall1987rheology}. Nevertheless, the wall's motion is stopped at the volume fraction of interest ($\phi=0.25-0.50$), and sufficient time is allowed for the network to relax to the nearest local equilibrium, with negligible net force and kinetic energy. All measurements were performed on ten independent, sufficiently relaxed, inherent states.
\section{Frictional gel simulation with non-central interaction forces}
Most prior simulation work, except for a few \cite {gadi2023micro,roy2016yieldinga,roy2016universality,roy2020micro} by the present author, focuses on colloidal gel systems with centrosymmetric potentials. However, in reality, in strong colloidal gels, strong van der Waals forces favor the rapid formation of solid-solid interparticle bonds that are resistant to rolling, sliding, and thermal fluctuations. Recent experiments\cite{bonacci2020contact,bonacci2022yield} also suggest a strong influence of the contact scale properties on the macroscopic response. On the issue of discontinuous shear thickening in the dense colloidal suspension, \cite{seto2013discontinuous,hsu2018roughness,mari2015discontinuous,lin2015hydrodynamic,singh2020shear}, it is established that the frictional interaction at the contact length scale is a necessary ingredient to reproduce the observed behavior. Therefore, the incorporation of non-hydrodynamic contact friction is sine qua non for accurately representing colloidal gel interactions at close distances and capturing the subsequent macroscopic response. Hence, we also simulate the colloidal gel system in the spirit of granular materials using the discrete element method (DEM) simulations.  

For an exact comparison with the depletion gel systems involving central interaction, all relevant parameters, such as system size, particle size, strength of thermal fluctuations, and compression strain rate, are kept identical to those used in the previous section. The only difference arises in the interparticle interaction. The contact forces are modeled according to the traditional DEM developed by Cundall and Strack \cite{cundall1979discrete}. Static friction is implemented by tracking the elastic portion of the shear displacement from the time of first contact. Particles $i$ and $j$, at positions ${\B r_i, \B r_j}$ with velocities ${\B v_i, \B v_j}$ and angular velocities ${\B \omega_i, \B \omega_j}$ experience a relative normal compression on contact given by $\Delta_{ij}=|r_{ij}-D_{ij}|$, where $r_{ij}$ is the center to center separation distance and $D_{ij}$ is the sum of the radii of the interacting pair; which gives rise to a  normal elastic repulsive force $ \B F^{(n)}_{ij} $. The normal force is modeled as a linear Hookean spring, and the tangential force is considered to vary linearly with the tangential displacement until the initiation of sliding. Normal and tangential components of the interaction force are given as:
\begin{eqnarray}
\B F^{(n)}_{ij}&=&k_n\Delta_{ij} \B n_{ij}-F_0\B n_{ij}-\frac{\gamma_n}{2} \B {v}_{n_{ij}}\\
\B F^{(t)}_{ij}&=&-k_t \B t_{ij}-\frac{\gamma_t}{2} \B {v}_{t_{ij}} 
\end{eqnarray}
Here $\Delta _{ij}$ and $t_{ij}$ represent normal and tangential displacements; $\B n_{ij}$ denotes the normal unit vector. $k_n$ and $k_t$ are the normal and tangential spring stiffness, respectively. $\gamma_n$ and $\gamma_t$ are the viscoelastic damping constants for the normal and tangential modes of deformation. $F_0$ is the magnitude of the constant attractive force acting center to center. This attractive force has zero range; it is activated only when a contact is made. $\B {v_n}_{ij}$ and $\B {v_t}_{ij}$ are respectively the normal and tangential components of the relative velocity between two particles. The relative normal and tangential velocities are given by
   \begin{eqnarray}
\B {v}_{n_{ij}}&=& (\B {v}_{ij} .\B n_{ij})\B n_{ij}  \\
\B {v}_{t_{ij}}&=& \B {v}_{ij}-\B {v}_{n_{ij}} - \frac{1}{2}(\B \omega_i + \B \omega_j)\times \B r_{ij}.
\end{eqnarray}
   where $\B {v}_{ij} = \B {v}_{i} - \B {v}_{j}$. Elastic tangential displacement $ \B t_{ij}$ is set to zero when the contact is first made and is calculated using $\frac{d \B t_{ij}}{d t}= \B {v}_{t_{ij}}$ and also the rigid body rotation around the contact point is accounted for to ensure that $ \B t_{ij}$ always remains in the local tangent plane of the contact \cite{silbert2001granular}.
The tangential force varies linearly with the relative tangential displacement at the contact point, as long as the tangential force does not exceed the limit set by the Coulomb limit
\begin{equation}
F^{(t)}_{ij} \le \mu F^{(n)}_{ij} \ , \label{Coulomb}
\end{equation}
where $\mu$ is the sliding friction coefficient. Note that the Coulomb inequality applies to the repulsive elastic normal force only. The rolling resistance between particles is also considered in the simulation, and it is modeled similarly to the contact elasticity and friction for the sliding mode. The rolling pseudo-force is given as \cite{luding2008cohesive},
\begin{equation}
\B F_{r}^{ij}=k_r\B \delta_{roll}-\gamma_r \B v_{roll}
\end{equation} 
 where $k_r$ denotes the rolling stiffness, $\B \delta_{roll}$ represents the rolling displacement, $\gamma_r$ is the damping constant for the rolling mode, $\B v_{roll}$ is the  relative rolling velocity. The rolling force is limited by $\mu_r F^{(n)}_{ij}$ similar to the sliding friction, where $\mu_r$ is the rolling friction coefficient. To make a one-to-one comparison with the depletion gel simulation, the stiffness ($k_n$) of the Hookean interaction is chosen in such a way that its value represents the corresponding bond stiffness value ($2\kappa^2U_0$) evaluated at the minimum of the Morse potential. The value of the constant attractive force at the contact, $F_0$, is taken as $U_0/\delta_c$, where $\delta_c$ denotes the range of attraction. Although in the frictional simulation the attractive force has a zero range, to connect the attractive force magnitude to the potential minimum used in the depletion simulation, a minimal attraction range, $\delta_c=0.03a$, is chosen. In this work, $k_n=k_t$, $k_r=k_t/4$, $\mu=\mu_r=0.5$.

Similar to the depletion gel system, the particles are subjected to a stochastic fluctuating force alongside a viscous drag force. Thermalization only happens to the translational degrees of freedom. The Langevin equation is solved to update the particle positions and velocities,
\begin{equation}
m_i \frac{d^2 \B x_i}{dt^2} =\B F_i -m_i\eta_t \frac{d\B x_i}{dt} +\B f_i(t) 
\end{equation}
where $\B f_i(t)$ is a $\delta$-correlated random force having zero mean with the following properties:
\begin{equation}
\langle \B f_i(t) \cdot \B f_j(t+\tau)\rangle = 2 \Gamma \delta (\tau) \delta_{ij}
\end{equation}
Here $\eta_t$ denotes the translational damping coefficient, $m_i$ is the mass of the particle $i$. $\B F_i$ is the net force on particle $i$ due to all pairwise interactions. $\Gamma$ is the strength of fluctuations and $\B x_i$ denotes the translation degrees of freedom. The rotational degrees of freedom are calculated from Newton's second law:
\begin{equation}
I_i \frac{d^2 \B \theta_i}{dt^2} =\B T_i
\end{equation}
where $I_i$ is mass moment of inertia of the particle $i$, $\B T_i$  is the net torque due to interparticle interactions, and $\B \theta_i$  denotes the rotational degrees of freedom.  

The global stress tensor is measured by taking averages of the dyadic products between the contact forces and the branch vector over all the contacts in a given volume.
  \begin{equation}
\sigma_{\alpha \beta} =\frac{1}{V}\sum_{j\neq i}\frac{r^{\alpha}_{ij} F^{\beta}_{ij} }{2}
  \end{equation}

\section{Snapshots of Gel and structure factor analysis}

\begin{figure}[htbp!]
\includegraphics[scale=0.14]{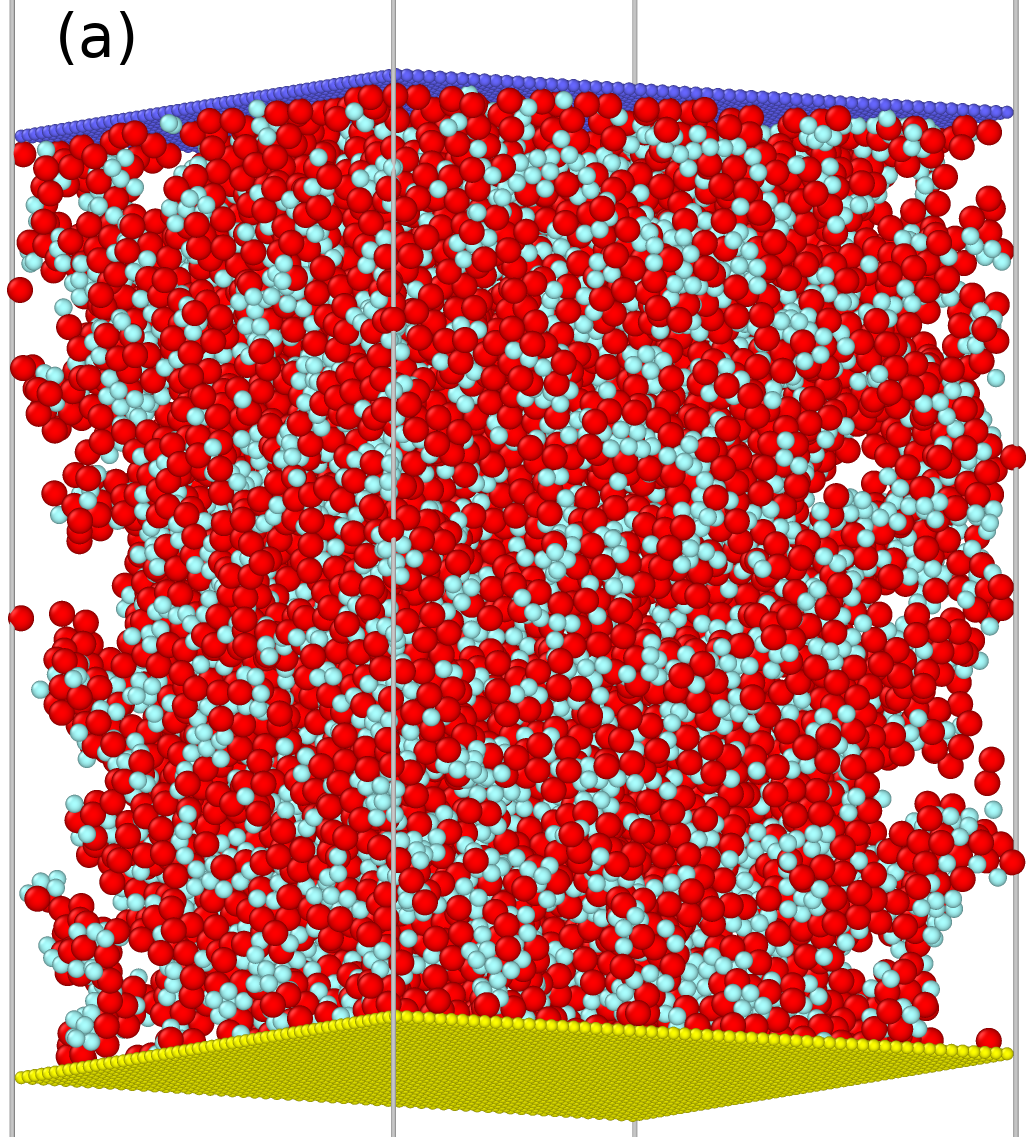}
\includegraphics[scale=0.14]{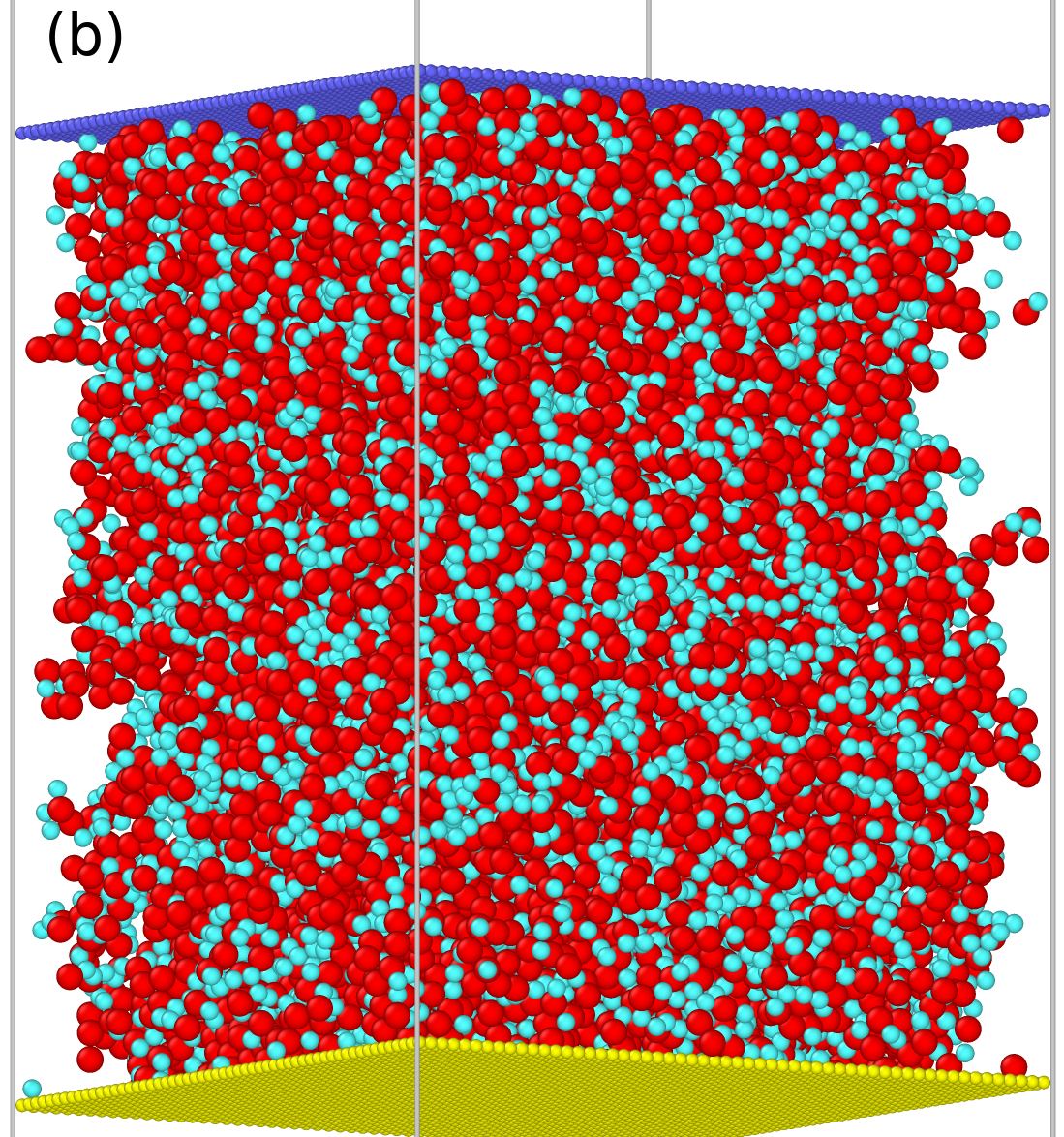}
\includegraphics[scale=0.16]{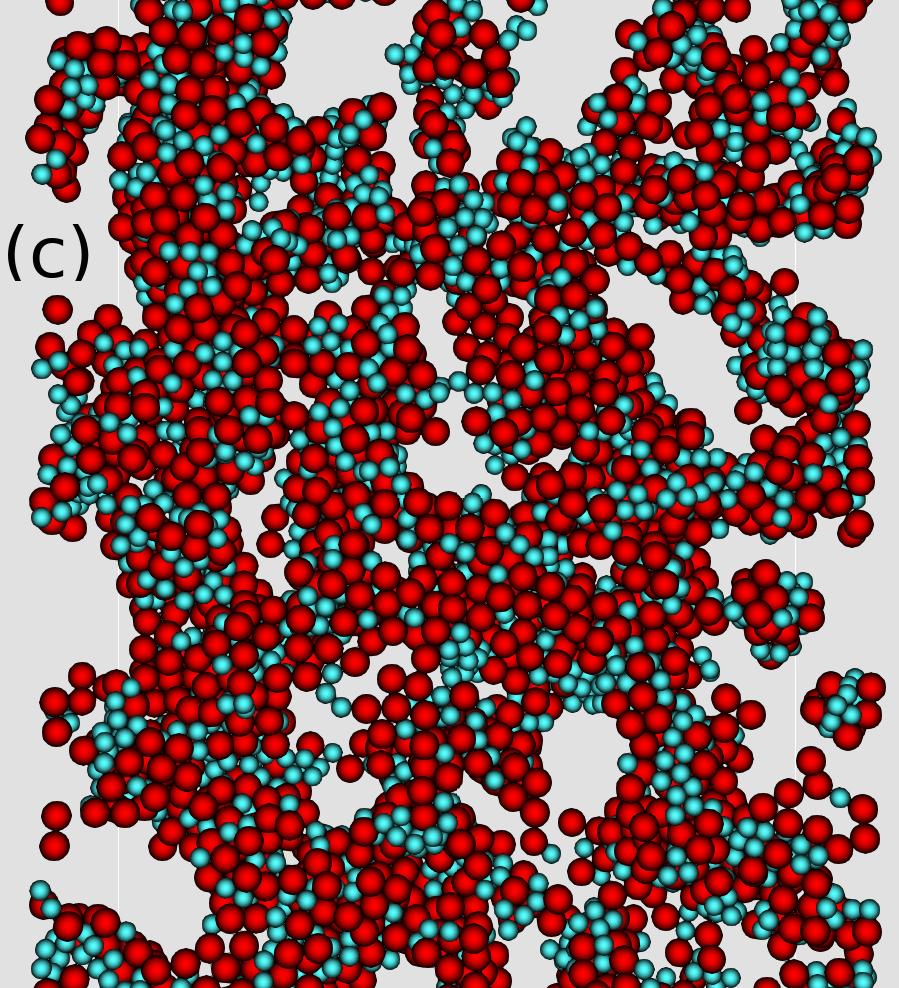}
\includegraphics[scale=0.16]{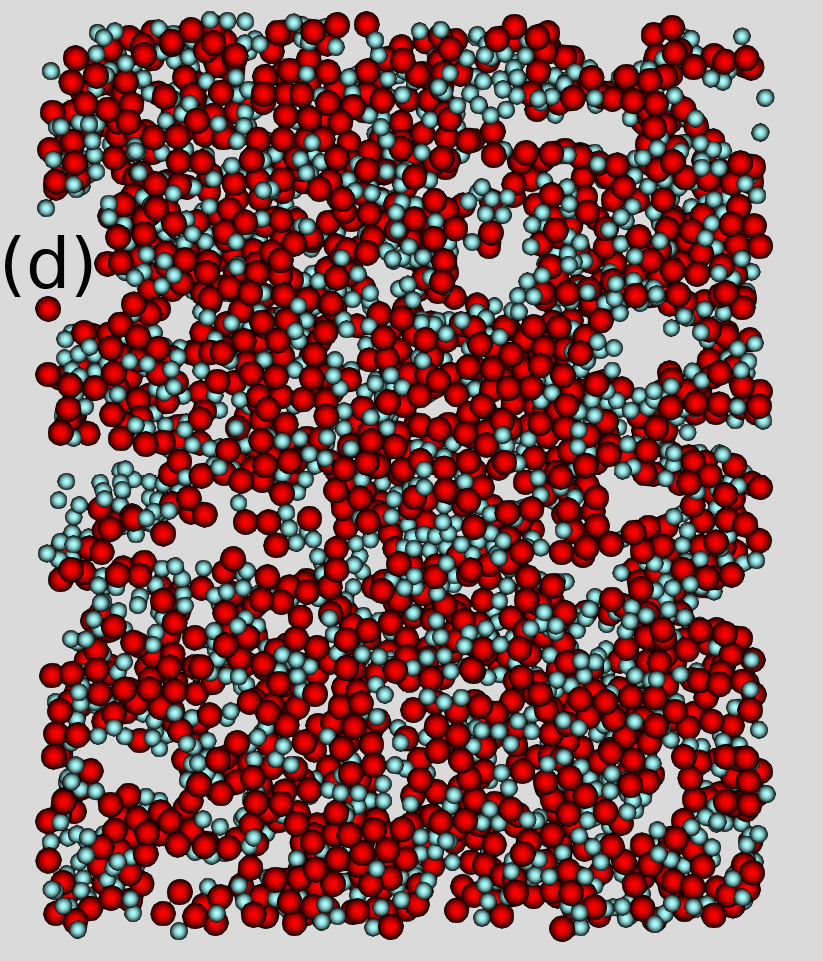}
\caption{(a) Illustration of typical depletion gel structure at $\phi=0.25$ for $\beta U_0=50$. (b) The gel network for the frictional case at the identical conditions. (c)-(d) shows the gel structure in a thin slice for the depletion and frictional gels, respectively.}
\label{SMFig1}
\end{figure}
\begin{figure}[htbp]
\includegraphics[scale=0.14]{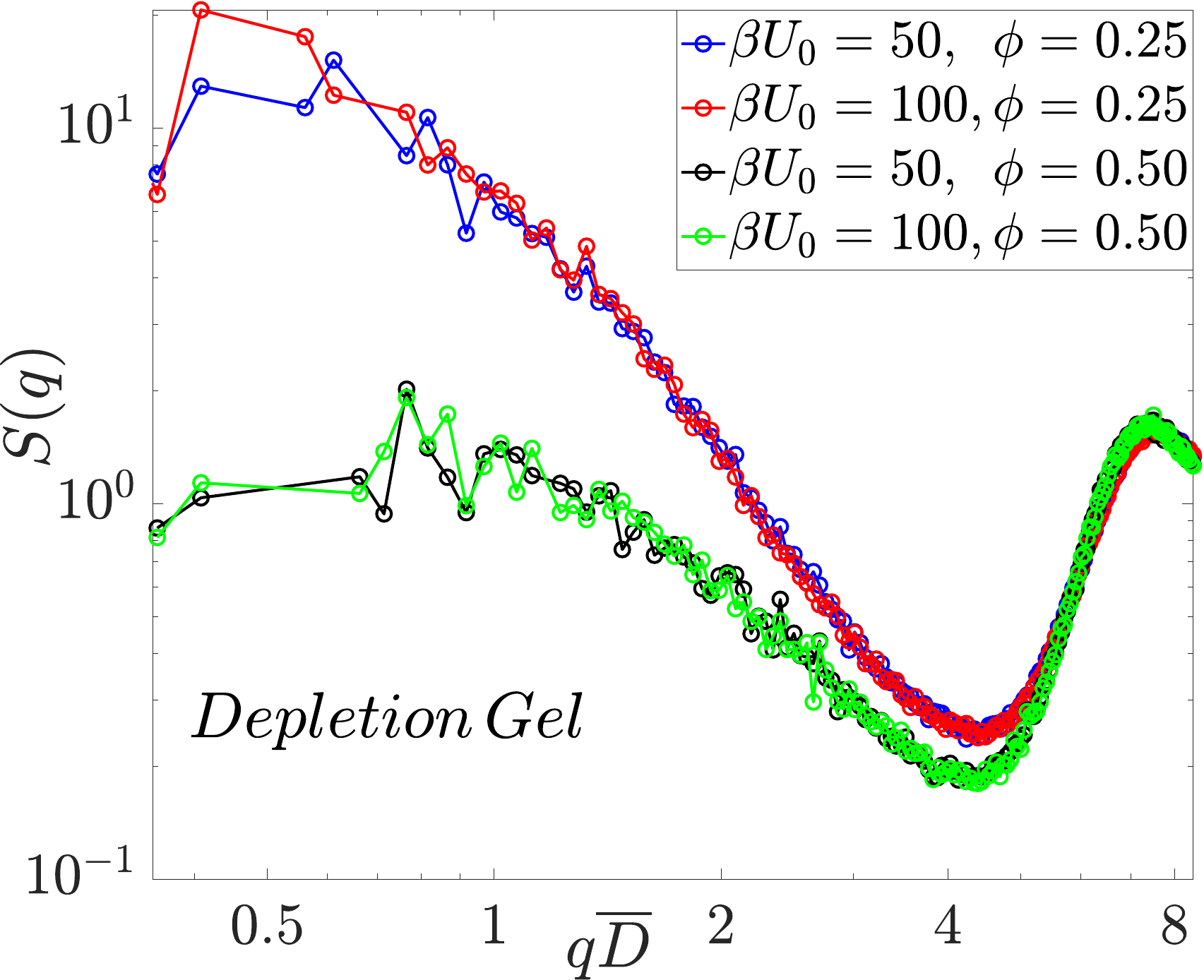}
\includegraphics[scale=0.143]{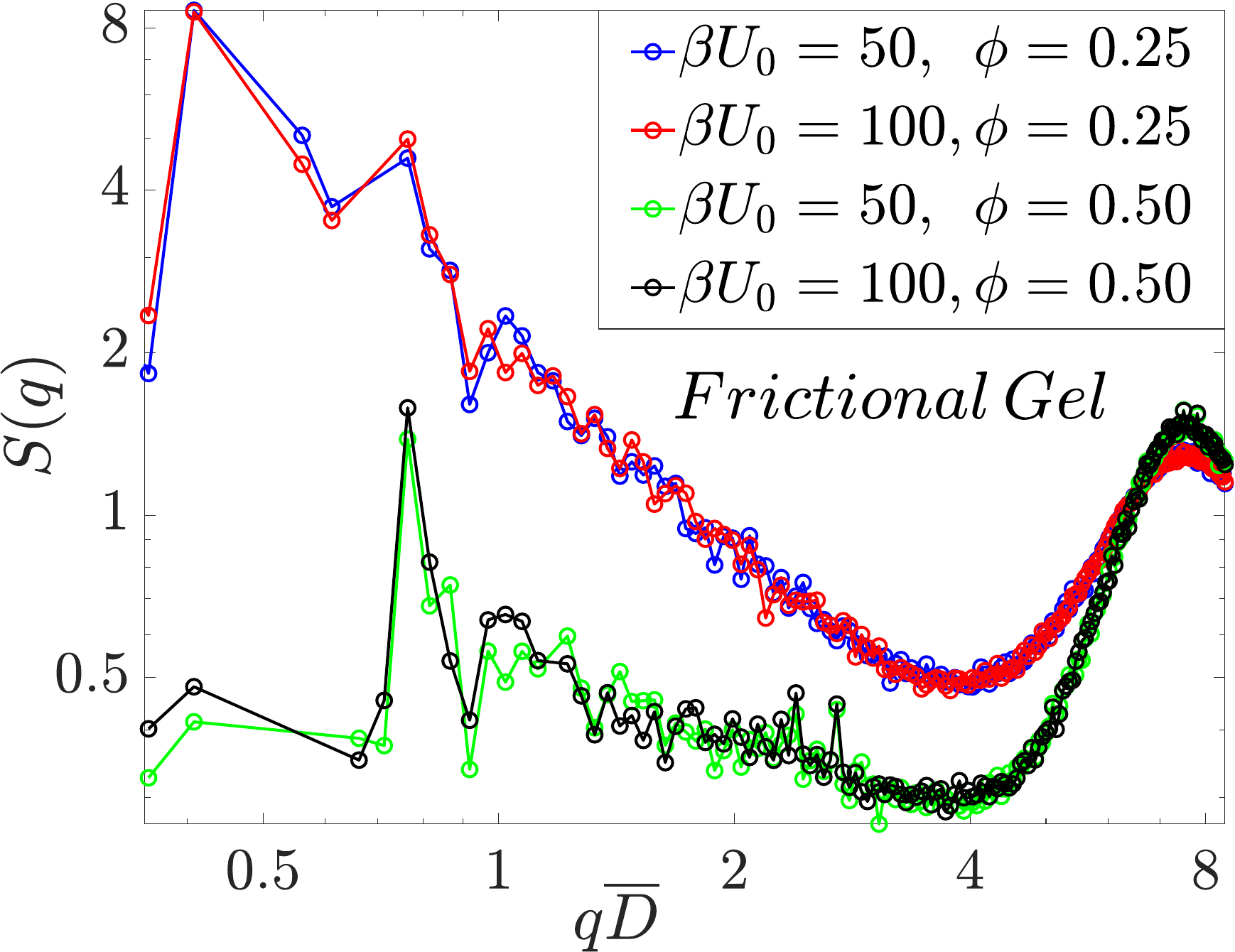}
\caption{(a) $S(q)$ is plotted for the depletion gel system for two different values of attraction strengths $\beta U_0=50$ and $100$ at the dilute and the dense regime. (b) $S(q)$ is plotted for the frictional gel at the identical conditions.}
\label{SMFig2} 
\end{figure}

In Fig.~\ref{SMFig1}, we show snapshots of the gel structure for both the depletion and frictional gels during the uniaxial consolidation process. The clusters formed in the depletion gel system are bulkier, and the connections between them are thicker than in gels with noncentral interactions. To further characterize the structure quantitatively, we evaluate the static structure factor by taking a direct Fourier transform of the particle positions and angularly averaging over all $\vec q$ of equal magnitude $q$ :
\begin{equation}
S(q)=\frac{1}{N}\left| \sum_{i=1}^{i=N} \exp\left(i\vec q\cdot \vec r_i\right)\right|^2
\end{equation}   
In Fig.~\ref{SMFig2}, we depict the structure factor for the two classes of gels for varied attraction strengths at different volume fractions. In all cases, the structure factor exhibits three notable features: (i) structural correlations are enhanced at the particle length scale, (ii) suppression of the correlations at the intermediate length scale, and (iii) increased structural correlations in the small wave number limit, which decreases with the increasing volume fractions.
The structure factor analysis reveals a key observation: the gel structure remains invariant across all length scales for the varied attraction strengths. 

\section{Isotropy in contact orientations} 
\begin{figure}[htbp]
\includegraphics[scale=0.34]{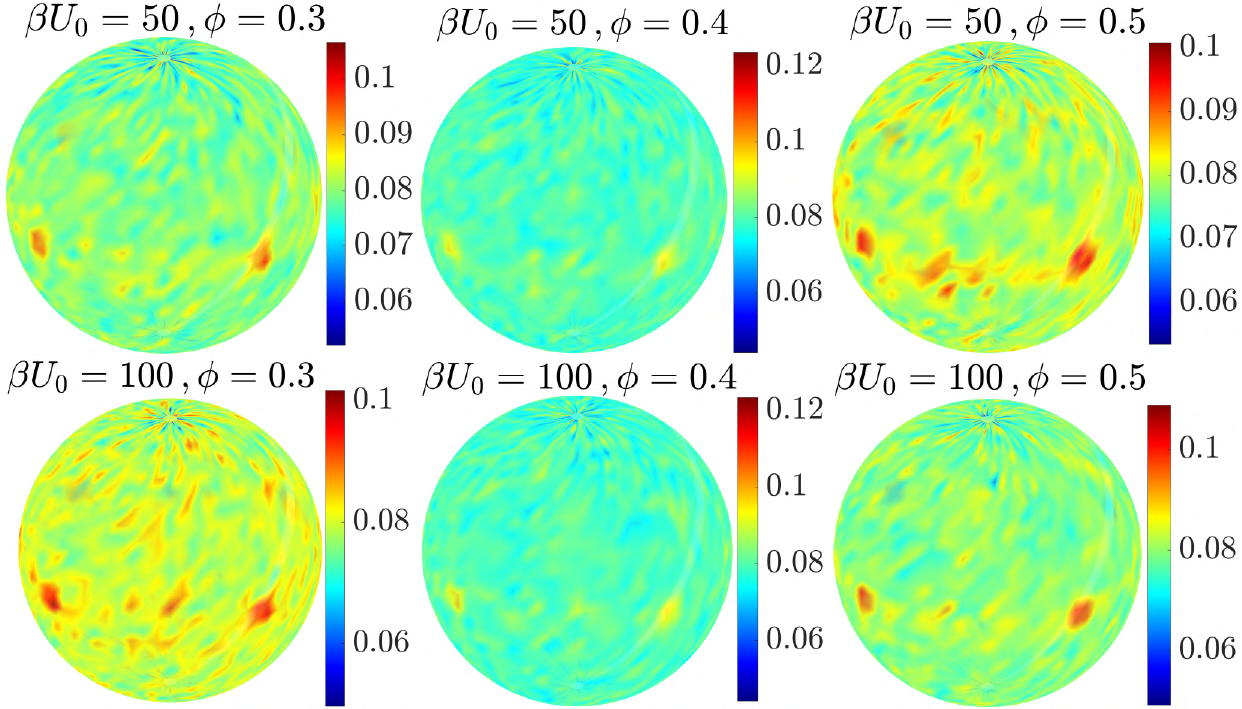}
\caption{Bond orientation probability density function, $P(\theta,\phi)$ for the depletion gel is plotted at different stages of the consolidation process and for different values of the potential minimum, $\beta U_0=50$ and $100$.}
\label{SMFig3} 
\end{figure}

\begin{figure}[htbp]
\includegraphics[scale=0.34]{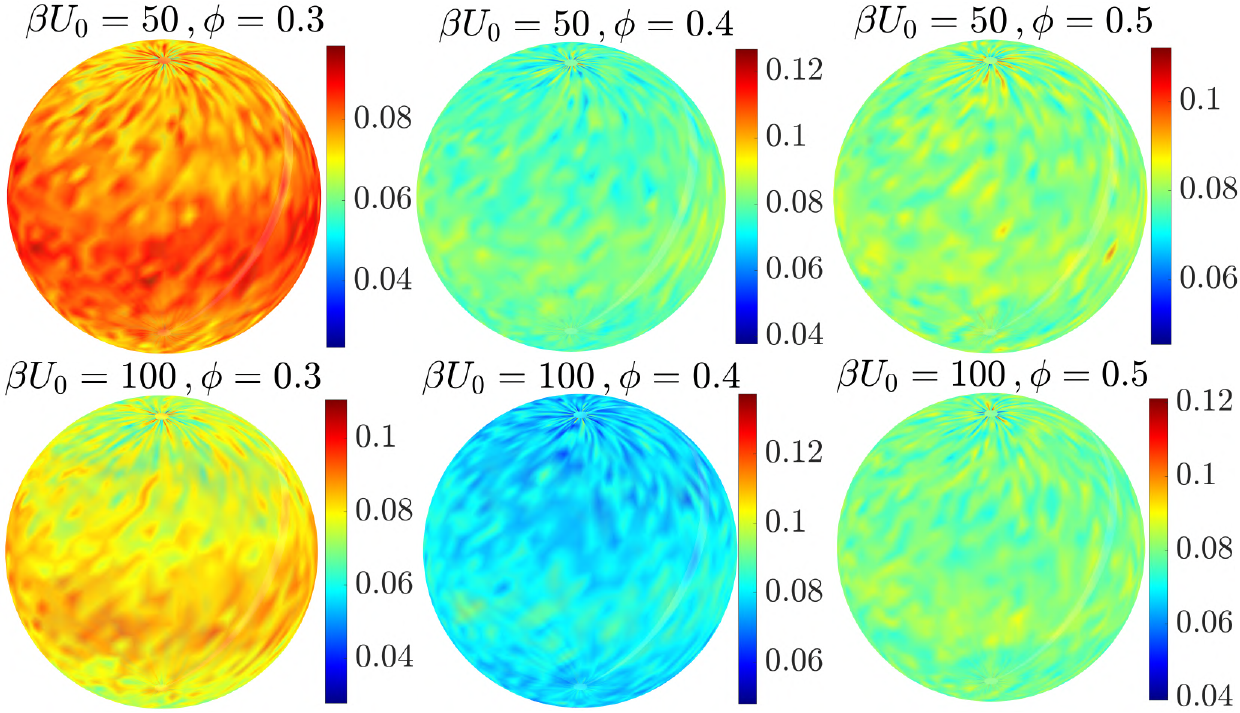}
\caption{Contact orientation probability density function, $P(\theta,\phi)$ for the frictional gel is plotted for different volume fractions and varied attraction strengths, $\beta U_0=50$ and $100$.}
\label{SMFig4} 
\end{figure}
Fig.~\ref {SMFig3} presents the probability density of interparticle bond orientations on the surface of a unit sphere for the depletion gel network. Intriguingly, the bond network remains isotropic for all volume fractions and attraction strengths. Similar isotropy in contact orientations is also observed for the gels with frictional interactions, as evidenced in Fig.~\ref{SMFig4}. Unlike dry granular materials, the persistence of isotropy in contact orientations for the gel network implies that contact anisotropy is not an essential ingredient for capturing the evolution of the macroscopic stress in the soft gel materials.

\section{Angular variations of pairwise interaction forces}
\begin{figure}[htbp]
\includegraphics[scale=0.43]{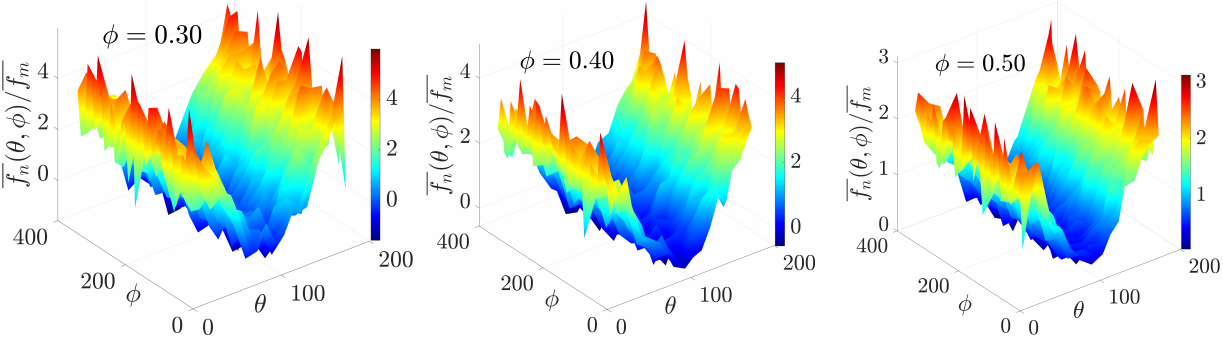}
\caption{Directional variation of the pairwise central interaction forces normalized by the mean normal force over all interactions is shown for the depletion gel system across different volume fractions. Here, $\beta U_0=50$.}
\label{SMFig5} 
\end{figure}

\begin{figure}[htbp]
\includegraphics[scale=0.43]{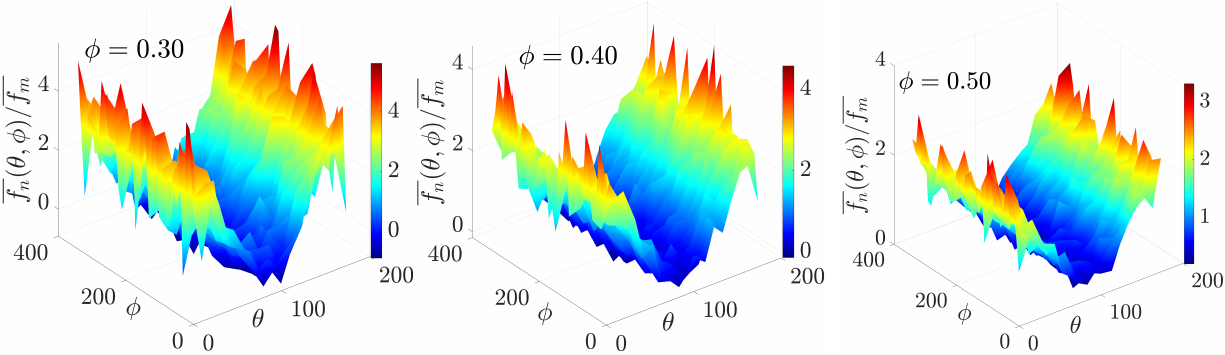}
\caption{Angular variation of the pairwise central interaction forces is shown for the depletion gel system across different volume fractions for $\beta U_0=100$.}
\label{SMFig6} 
\end{figure}

In Fig.~\ref{SMFig5}, we show the angular modulation of the center-to-center interaction force, $\overline{f_n}(\theta,\phi)$, for the depletion gel system at various stages of the uniaxial consolidation process for $\beta U_0=50$. The same is also plotted for a higher attraction strength $\beta U_0=100$ in Fig.~\ref{SMFig6}. The universal structure of angular modulation in interaction forces across different attraction strengths and volume fractions is appealing. It conforms to the form as predicted in the main text.

\begin{figure}[htbp]
\begin{center}
\includegraphics[scale=0.33]{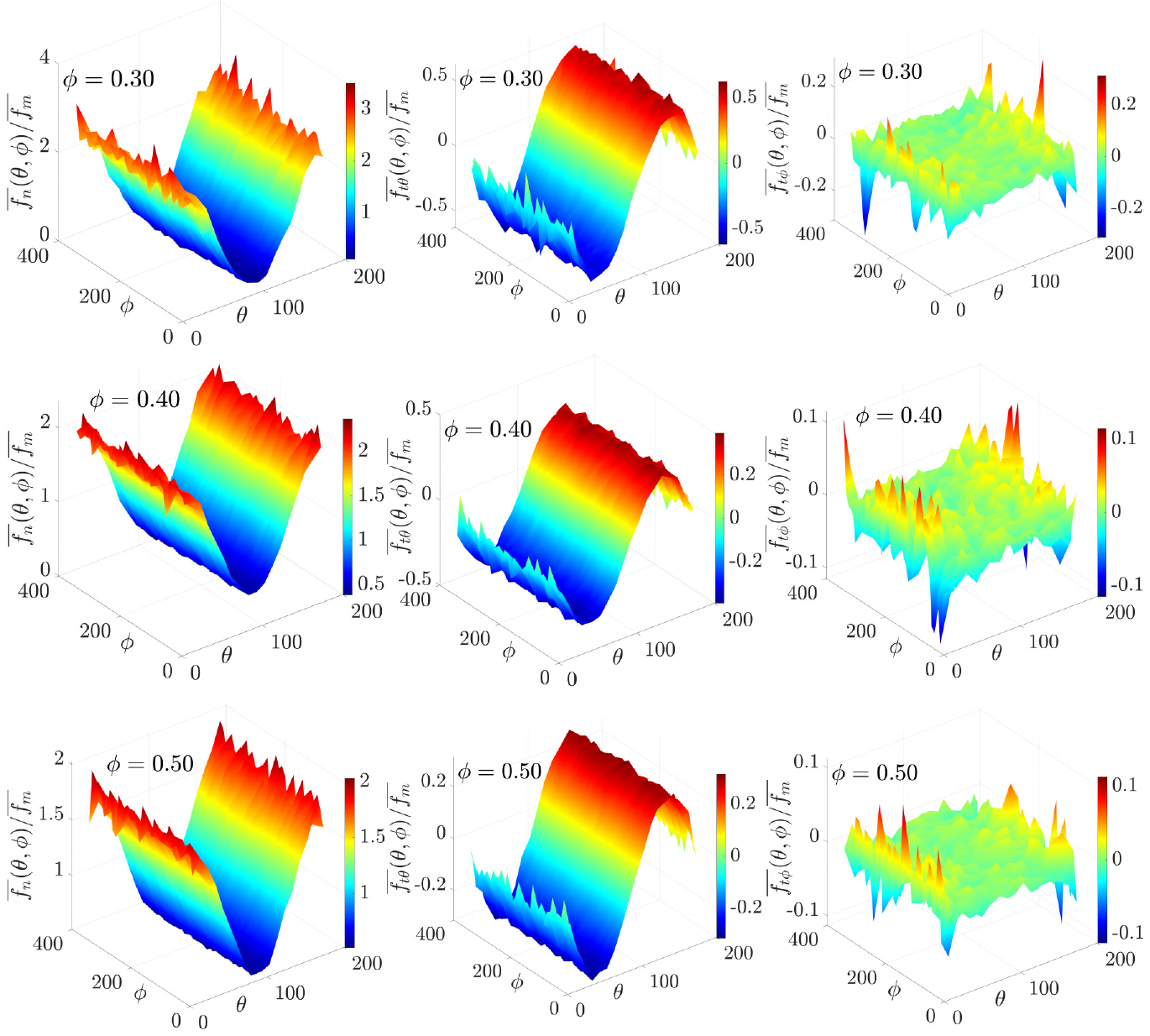}
\caption{Directional variation of the different components ($\overline{f_n}(\theta,\phi)$, $\overline{f_{t\theta}}(\theta,\phi)$, $\overline{f_{t\phi}}(\theta,\phi)$) of the pairwise interaction force normalized by the mean normal force over all interactions is shown for the frictional gel system across different volume fractions. Here, $\beta U_0=50$.}
\label{SMFig7} 
\end{center}
\end{figure}

\begin{figure}[htbp]
\begin{center}
\includegraphics[scale=0.33]{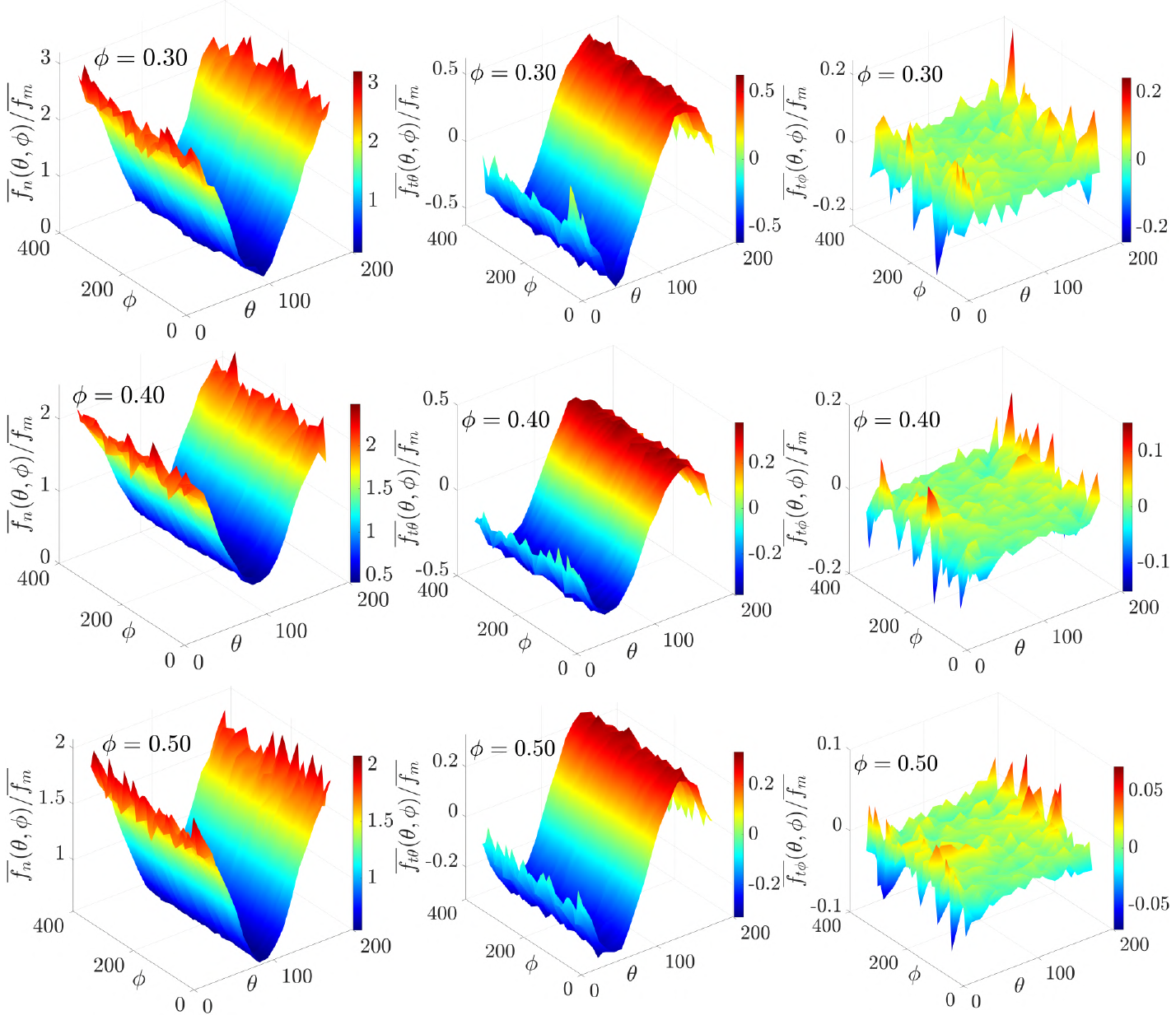}
\caption{Angular variation of $\overline{f_n}(\theta,\phi)$, $\overline{f_{t\theta}}(\theta,\phi)$, $\overline{f_{t\phi}}(\theta,\phi)$ as observed for the frictional gel system across different volume fractions for $\beta U_0=100$.}
\label{SMFig8} 
\end{center}
\end{figure}

The noncentral pairwise interaction in the frictional gel system will give rise to polar and azimuthal components in addition to the center-to-center normal component. The azimuthal component is expected to be negligible due to the absence of contact scale twisting friction and the externally imposed rotational field. Hence, the tangential force will mainly consist of the polar component. The angular modulation of the normal, polar, and azimuthal components of the pairwise interaction force is shown for the frictional gel force network for different volume fractions and attraction strengths in Fig.\ref{SMFig7}-\ref{SMFig8}. Again, the orientational structure of the components of the interaction force exhibits a universal form, as dictated by the model in the main manuscript. 

\bibliography{gel}

\end{document}